\documentclass[aps,notitlepage,12pt,prl,superscriptaddress,a4paper,floatfix,showpacs]{revtex4-1}

\usepackage{graphicx,color}
\usepackage{amsmath,amssymb}
\usepackage{verbatim}
\usepackage{ulem}
\usepackage{mathptmx}

\usepackage{pifont}

\long\def\symbolfootnote[#1]#2{\begingroup%
\def\thefootnote{\fnsymbol{footnote}}\footnote[#1]{#2}\endgroup} 
\def\blfootnote{\xdef\@thefnmark{}\@footnotetext}

\newcommand{\be}{\begin{equation}}
\newcommand{\ee}{\end{equation}}
\newcommand{\bi}{\begin{itemize}}
\newcommand{\ei}{\end{itemize}}
\newcommand{\bea}{\begin{eqnarray}}
\newcommand{\eea}{\end{eqnarray}}

\begin{document}

\title{Ultra-bright GeV photon source via controlled electromagnetic cascades in laser-dipole waves} 
\author{A.~Gonoskov}
\affiliation{Department of Physics, Chalmers University of Technology, SE-41296 Gothenburg, Sweden}
\affiliation{Institute of Applied Physics, Russian Academy of Sciences, Nizhny Novgorod 603950, Russia}
\affiliation{Lobachevsky State University of Nizhni Novgorod, Nizhny Novgorod 603950, Russia}
\author{A.~Bashinov}
\affiliation{Institute of Applied Physics, Russian Academy of Sciences, Nizhny Novgorod 603950, Russia}
\author{S.~Bastrakov}
\affiliation{Lobachevsky State University of Nizhni Novgorod, Nizhny Novgorod 603950, Russia}
\author{E.~Efimenko}
\affiliation{Institute of Applied Physics, Russian Academy of Sciences, Nizhny Novgorod 603950, Russia}
\author{A.~Ilderton}
\affiliation{Department of Physics, Chalmers University of Technology, SE-41296 Gothenburg, Sweden}
\affiliation{Centre for Mathematical Sciences, University of Plymouth, PL4 8AA, UK}
\author{A.~Kim}
\affiliation{Institute of Applied Physics, Russian Academy of Sciences, Nizhny Novgorod 603950, Russia}
\author{M.~Marklund}
\affiliation{Department of Physics, Chalmers University of Technology, SE-41296 Gothenburg, Sweden}
\author{I.~Meyerov}
\affiliation{Lobachevsky State University of Nizhni Novgorod, Nizhny Novgorod 603950, Russia}
\author{A.~Muraviev}
\affiliation{Institute of Applied Physics, Russian Academy of Sciences, Nizhny Novgorod 603950, Russia}
\author{A.~Sergeev}
\affiliation{Institute of Applied Physics, Russian Academy of Sciences, Nizhny Novgorod 603950, Russia}

\maketitle

One aim of upcoming high-intensity laser facilities~\cite{eli, xcels, vulcan} is to provide new high-flux gamma-ray sources~\cite{eli_secondary_sources}. Electromagnetic cascades~\cite{Bell2008, Bulanov20102,Fedotov2010,Nerush2011, tamburini.arxiv.2016} may serve for this, but are known to limit both field strengths and particle energies~\cite{RDR}, restricting efficient production of photons to sub-GeV energies~\cite{ridgers.pop.2013, jirka.pre.2016, grismayer.pop.2016}. Here we show how to create a directed GeV photon source, enabled by a controlled interplay between the cascade and anomalous radiative trapping~\cite{ART}. Using advanced 3D QED particle-in-cell (PIC) simulations~\cite{gonoskov.pre.2015} and analytic estimates, we show that the concept is feasible for planned~\cite{vulcan} peak powers of 10~PW level. A higher peak power of 40~PW can provide $10^9$ photons with GeV energies in a well-collimated 3~fs beam, achieving peak brilliance ${9 \times 10^{24}}$~ph~s$^{-1}$mrad$^{-2}$mm$^{-2}$/0.1${\%}$BW. Such a source would be a powerful tool for studying fundamental electromagnetic~\cite{DiPiazza:2011tq} and nuclear processes~\cite{eli,Spring8,MAMI}.

Advances in high-intensity laser science offers opportunities for creating a new kind of high flux gamma-ray source, based on the use of strong laser fields to accelerate particles and stimulate emission within a single optical cycle~\cite{ridgers.pop.2013, jirka.pre.2016, grismayer.pop.2016, vranic.arxiv.2016, gong.arxiv.2016}. However, from a naive consideration of particle dynamics one would expect particles to be expelled from the temporal and spatial regions which are optimal for energy gain (i.e.~the electric field antinodes). Furthermore, for intensities above~$10^{24}$~W/cm$^2$ radiation losses prevent particles from reaching their potential maximum energy (during a single phase of acceleration). This limits the effective generation of photons to sub-GeV energies.~\cite{RDR,Yu.SciRep.2016}

\begin{figure}[t!!]
	\centering\includegraphics[width=\columnwidth]{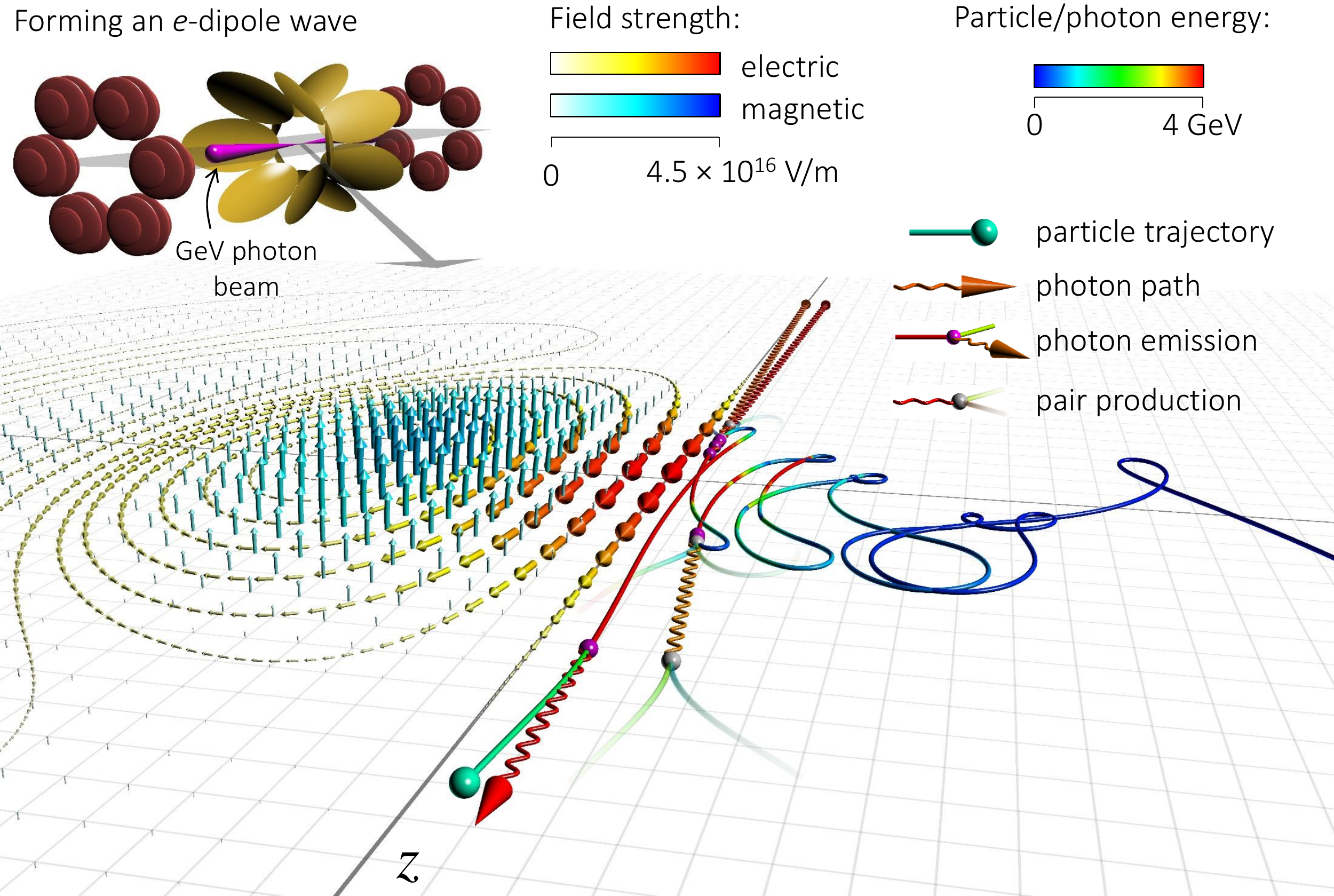}
	\caption{\label{fig:Fig1} A schematic illustration of the concept of anomalous radiative trapping and photon emission. In the top left we show how to generate a dipole wave using a set of off-axis parabolic mirrors (yellow) and 12 laser pulses (dark red). The field structure of the dipole wave is shown in the left half of the horizontal plane (there is axial symmetry about the $z$-axis). We also show a trajectory of one of the seeding electrons taken from a simulation with $P=200$~PW. Photons of energy above 3~GeV are shown together with their emission events (magenta spheres). While some of these photons contribute to cascade development (grey spheres), others form a brilliant source emitted along the z-axis, also shown in the top left.}
\end{figure}

Our aim here is to find the optimal strategy for source creation. To do so we exploit the anomalous radiative trapping~\cite{ART} (ART) of electrons and positrons in a dipole wave~\cite{gonoskov.pra.2012}, the latter being the field configuration which provides the highest possible field strength for a given peak power of radiation. The dipole wave can be formed by orienting, timing and focusing a number of laser pulses~\cite{ART} as shown in Fig.~\ref{fig:Fig1}. In this configuration, ART traps particles and makes them oscillate in spatiotemporal regions ideal for gaining a maximal possible energy. Furthermore, within each oscillation of the field the particles have a high probability of emitting a large portion of their gained energy in the form of a single photon. However, the dominance of this favourable single-particle dynamics requires a total power of around 100~PW. As we will see further, even for a lower power a significant number of the generated high-energy photons can decay into $e^+e^-$ pairs giving rise to a cascade of pair production. As a result, before the peak intensity of the field is reached the cascade can generate an $e^+e^-$ plasma dense enough to not only restrict the growth of the field intensity~\cite{Fedotov2010}, but also to terminate the favourable ART-regime dynamics. By assessing these processes with advanced simulations, we reveal here that for powers of the 10~PW level a new physical scenario arises in which the cascade and ART in fact induce and support each other. We demonstrate that controlling the cascade development by matching the laser pulse intensity and duration in realistic ranges makes it possible to create a unique and ultra-efficient photon source in which laser radiation is converted into a well-collimated flash of GeV photons with unprecedented brightness.

The key to our concept is matching the pulse duration, peak power and initial particle density in the target, such that the maximal field intensity is reached just before collective plasma effects start to cause a significant reduction in the energy of generated photons.This requires that particle growth in the leading half of the laser pulse be restricted, such that the number of particles remains below a certain critical value $N_{\text{max}}$ which would cause significant back-reaction on the field. By ensuring that this constraint is fulfilled, we can arrange for a maximal number of particles to interact with the most intense part of the laser pulses, and emit a large number of high-energy photons, as desired. To perform a theoretical analysis and thus obtain estimates for the above constraint we begin by describing how particle dynamics in the ART regime~\cite{ART} governs the cascade.

The motion of particles in the ART regime, illustrated in Fig.~\ref{fig:Fig1}, is quasi-periodic and consists of two qualitatively different phases.  During the half-cycle when the electric field does not change sign, a trapped particle is accelerated and gains a kinetic energy of around $amc^2$ before starting to emit frequently (hereafter $a \approx 800 \sqrt{P/\left(1 \text{PW}\right)}$ is the electric field amplitude in relativistic units for a dipole wave of total incoming power $P$); we refer to this as the \textit{acceleration phase}. As the magnetic field strength rises the particle loses a majority of its energy, having possibility to emit photons with energies of up to $amc^2$, and then starts gyrating; we refer to this as the \textit{turning phase}, and this lasts until the electric field of the next acceleration phase starts to force the particle back in the opposite direction. Due to the probabilistic nature of emission in quantum electrodynamics, it may happen that the particles do not experience sufficient losses to be forced back, and hence leave the trapped state as shown in Fig.~\ref{fig:Fig1}.

Using estimates for the magnetic field strength in dipole waves~\cite{gonoskov.pra.2012}, one can obtain that for powers of order $P = 10$~PW the quantum efficiency parameter~\cite{Ritus,DiPiazza:2011tq}~$\chi$ attains values of order unity. This parameter is defined as the ratio $\chi=\gamma F_{\rm{eff}}/(eE_S)$ of the force acting perpendicular to the direction of motion of the electron in its rest frame, to that force which it would experience in a field of Sauter-Schwinger strength $E_S=m^2c^3/e\hbar\sim 10^{16}$V/cm ($e$ and $\gamma$ are the electron charge and gamma factor respectively). The fact that $\chi\gtrsim 1$ means that a significant part of an electron's energy can emitted as a single photon, which can then convert into an electron-positron ($ee^+$) pair. By equating the time needed for this to happen with the duration of the turning phase we can (see Methods B) estimate that this process gives rise to a pair production cascade for powers $P>P_\text{min} \sim 15$~PW. As we will see below, this analytically determined threshold value is in a good agreement with numerical results. However, as we demonstrate in Methods A, cascade evolution in our system is a very complex process which is not amenable to simple analytical estimates. Thus in the interests of providing accurate predictions we turn to simulations.

We use the 3D PIC code PICADOR\cite{bastrakov.jcp.2012} with quantum extensions~\cite{gonoskov.pre.2015} (see Methods C) to simulate cascade development in a dipole wave of constant incoming power~$P$, for several values of $P$. We calculate the growth rate $\Gamma$ of particle number averaged over a half-period, using the relative increase of the number of trapped particles during a single turning phase. This takes into account both particle production and particles leaving the trapping state.

\begin{figure}[t!!]
	\centering\includegraphics[width=\columnwidth]{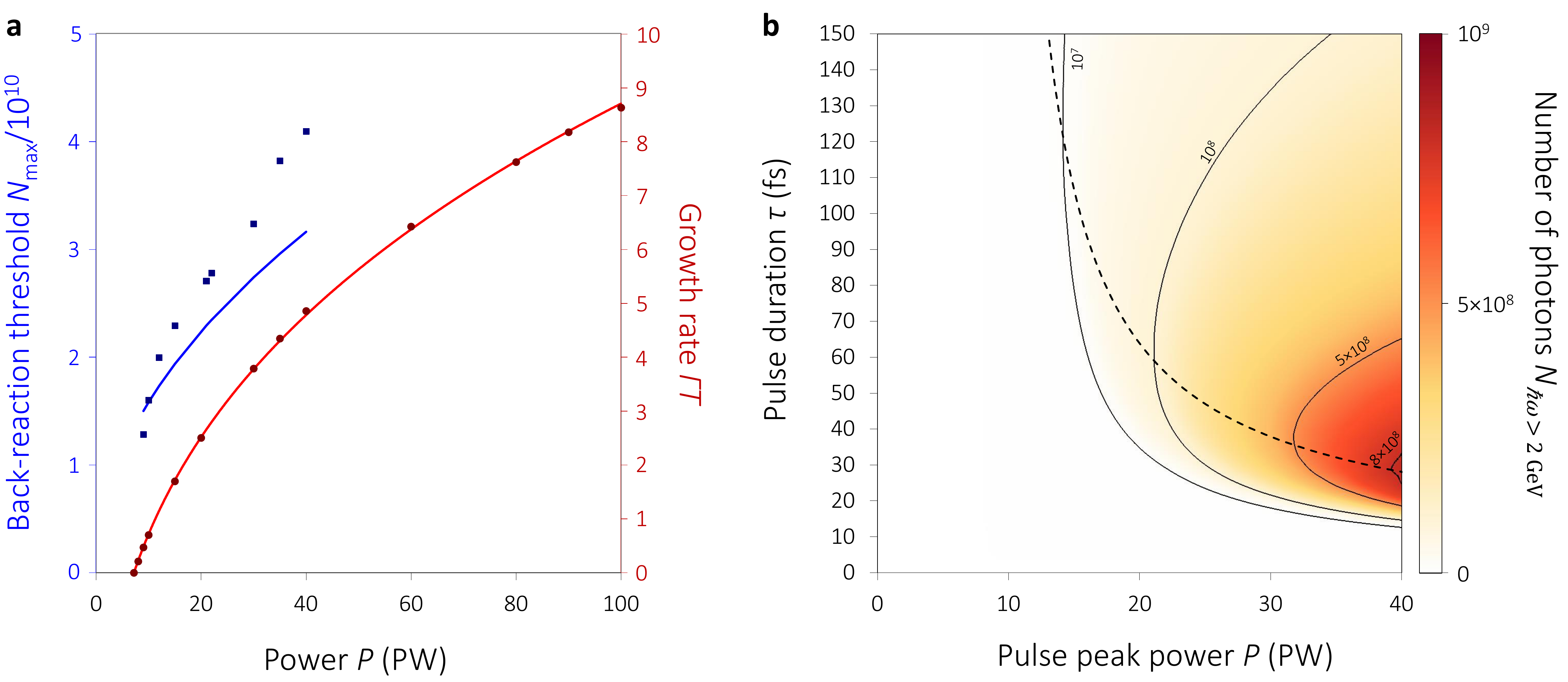}
	\caption{\label{fig:Fig2} Simulation results: a.) rate of particle population growth (red, right hand scale), and the dependency of the back-reaction threshold  (blue, left hand scale), both as a function of power $P$. The analytical estimate for the back-reaction threshold is shown with a solid blue line, and agrees very well with the numerical results shown as blue points. The accurate numerical approximation to the data points obtained from simulation is shown with a solid red curve, and this is used to derive the estimate (\ref{est1}), shown in panel b): the expected number of photons with energy above 2~GeV, as calculated based on numerical studies (see methods), when using a Gaussian pulse of total peak power $P$ and duration $\tau$. The analytically obtained optimal condition (\ref{est1}) is shown with the black dashed curve: this clearly runs through, the centre of the optimal region.}
\end{figure}

On time scales longer than a half-period, the particle number grows exponentially before back-reaction sets in. For this exponential stage the rate $\Gamma$ depends only on $P$ and, according to the  results shown in Fig.~\ref{fig:Fig2}~(a), may be approximately fitted to the curve $\Gamma(P) \approx 3.21 T^{-1} \big([P/(1~\mathrm{PW})]^{1/3} - [P_\text{min}/(1~\mathrm{PW})]^{1/3})$, where $T$ is the laser period and $P_\text{min} \simeq 7.2$~PW is the numerically determined threshold value required for the number of particles to grow. We integrate the pair production rate over the leading half of a laser pulse with Gaussian profile, FWHM duration $\tau$, and demand that by the instant of reaching the peak power the number of trapped particles rises from $N_0 = 10$ (taken to guarantee the cascade seeding) to $N_{\text{max}}$, introduced above. By equating the Coulomb field of the trapped particles to $a/2$, i.e.~by demanding no more than $50\%$ collective effects (see agreement with numerical results in Fig.~\ref{fig:Fig2}~(a), solid blue curve) we estimate
\be
	N_{\text{max}} \approx 10^9\sqrt{P} \;,
\ee
and obtain the following optimal condition on the pulse duration $\tau$, at $P>P_\text{min}$:
\begin{equation}\label{est1}
	\frac{\tau}{T} \simeq \frac{2.23 \log[5\times10^8 P^{1/2}] } {P^{0.59}-3.6} \;.
\end{equation}
This curve is shown in Fig.~\ref{fig:Fig2}~(b) and fits well with the numerically estimated (see methods) number of produced photons with energy $\hbar \omega > 2$~GeV. As can be seen from the Figure, a large number of energetic photons are produced and escape the fields, despite a rapid cascade process. This is because their mean free path is roughly four times larger than that of electrons/positrons with the same energy, and in addition is proportional to $\left(\hbar \omega \right)^{1/3}$, which favours the escape of photons with higher energies.

\begin{figure}[t!!]
	\centering\includegraphics[width=0.6\columnwidth]{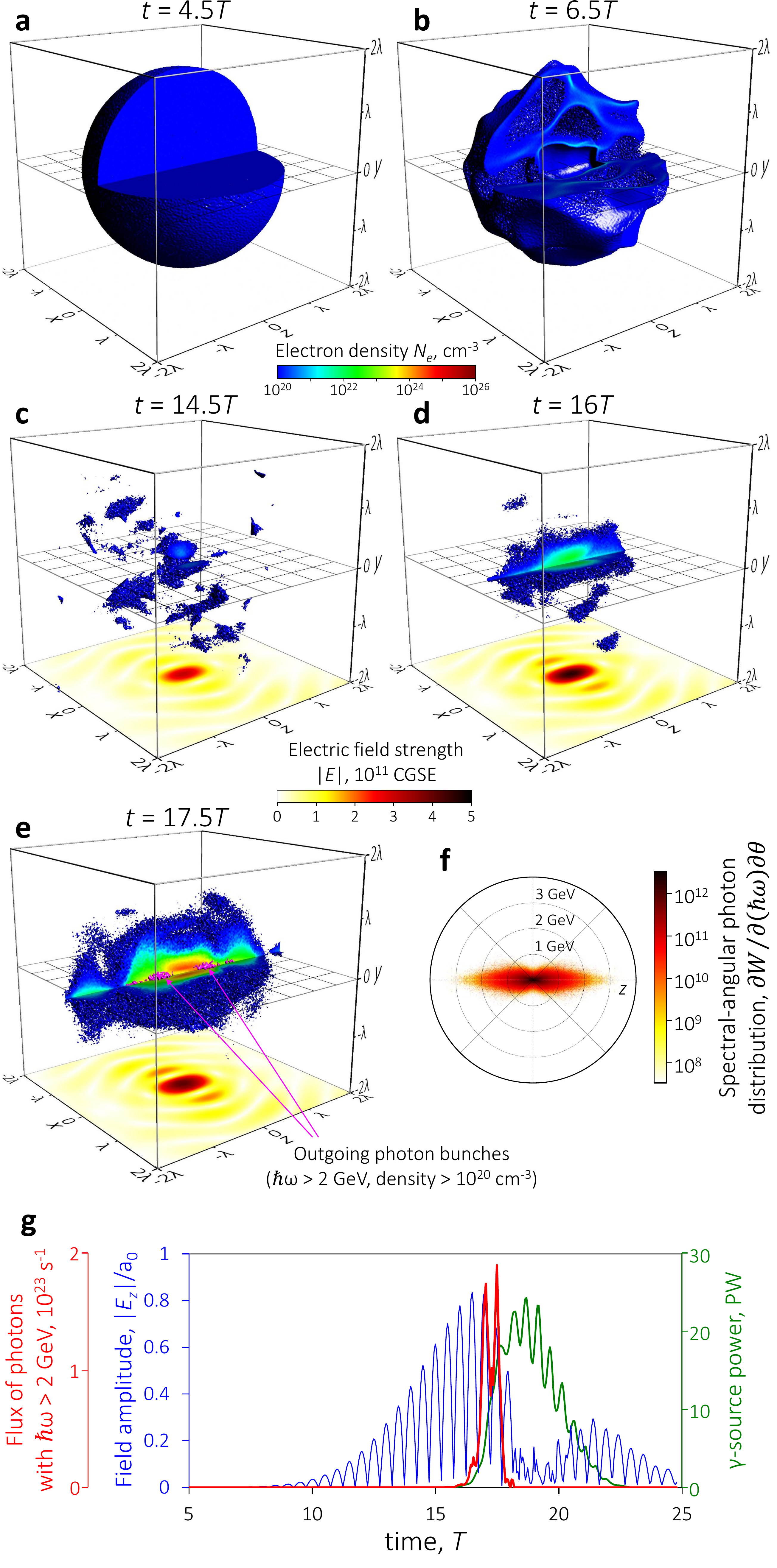}
	\caption{\label{fig:Fig3} The results of 3D PIC simulations for a dipole-wave formed by 12 slightly de-phased laser pulses of 15~fs duration (FWHM for intensity) and total peak power of 40~PW. The electron density~$>10^{20}$~cm$^{-3}$ and electric field strength in the plane $y = 0$ (shifted downward) are shown for several instances: 
	(a) the initial target,
	(b) the target affected by the leading edge of the pulses, 
	(c) ART beginning to trap particles and start the cascade,
	(d) the cascade development, 
	and (e) the generation of GeV photons. The overall angular-resolved spectrum of emitted photons is shown in panel (f). Panel (g) shows the temporal evolution of the field strength in the centre (blue), the total power of generated photons (green) and the flux of photons with energy $>$~2~GeV (red).} 
\end{figure}   

Having now identified the required pulse parameters, we provide a concrete example of the possible source properties using a comprehensive PIC simulation. (For full simulation details see Methods.) The results are shown in Fig.~\ref{fig:Fig3} (see also Supplementary Video). We form the dipole wave using 12 laser pulses with a total peak power of 40~PW and each having $15$~fs duration (FWHM for intensity), which is matched to the initial particle density of the target (a uniform sphere with diameter 3~$\mu$m and density $10^{20}$~cm$^{-3}$). Instead of the considered target, in reality adjusting density of stray particles in the chamber can be used to trigger the cascade starting from an arbitrarily small number of particles. (This is because the cascade can be even prevented by choosing high enough vacuum \cite{Gonoskov:2013ada, tamburini.arxiv.2016}.) To account for realistic experimental conditions, and so demonstrate the robustness of our concept, the pulses are slightly de-phased (randomly within $1/30$ of the wave period). The simulation shows that the concept provides $10^9$ photons with energies $\hbar \omega > 2$~GeV in a well collimated beam ($ < 5^{\circ}$) and of just 3~fs duration, achieving peak flux of $1.9 \times 10^{23}$~ph~s$^{-1}$, brightness $7.4 \times 10^{26}$~ph~s$^{-1}\text{mrad}^{-2}\text{mm}^{-2}$ and brilliance $9 \times 10^{24}$~ph~s$^{-1}$mrad$^{-2}$mm$^{-2}$/0.1$\%$BW, exceeding peak values of laser Compton scattering sources in GeV range by several orders of magnitude \cite{weller.ppnp.2009}. Such a source with extraordinary high flux of GeV photons with a broad spectrum will open qualitatively new possibilities for studying photo-nuclear processes\cite{eli,Spring8,MAMI}, such as triggering multiple transitions between short-living states. One could also take advantage of the even higher photon flux in the focal region, by placing nuclear targets in or as close as possible to the focal spot.
 
\clearpage

\section*{Methods}

\subsection{Particle dynamics and the cascade}

Dynamics and radiation losses in the ART regime are highly nonlinear, as even qualitative explanations assuming continuous (classical) radiation losses show~\cite{ART}. The discrete and stochastic nature of quantum emission does not destroy ART, but instead adds another layer of complexity. To analyse typical particle motion in the ART regime we therefore use here a `regularised' model of emission which is a course-graining of the full quantum model (as based on the locally constant field approximation and implemented in PIC codes). In the model, particles move exactly along their mean free paths. During this motion we integrate the probability of emission until it reaches unity, and then a photon is emitted with an energy equal to the average of that which it would have according to the QED rates. Local values of $\chi$ and $\gamma$ are used throughout.

\begin{figure}[t!!]
	\centering\includegraphics[width=\columnwidth]{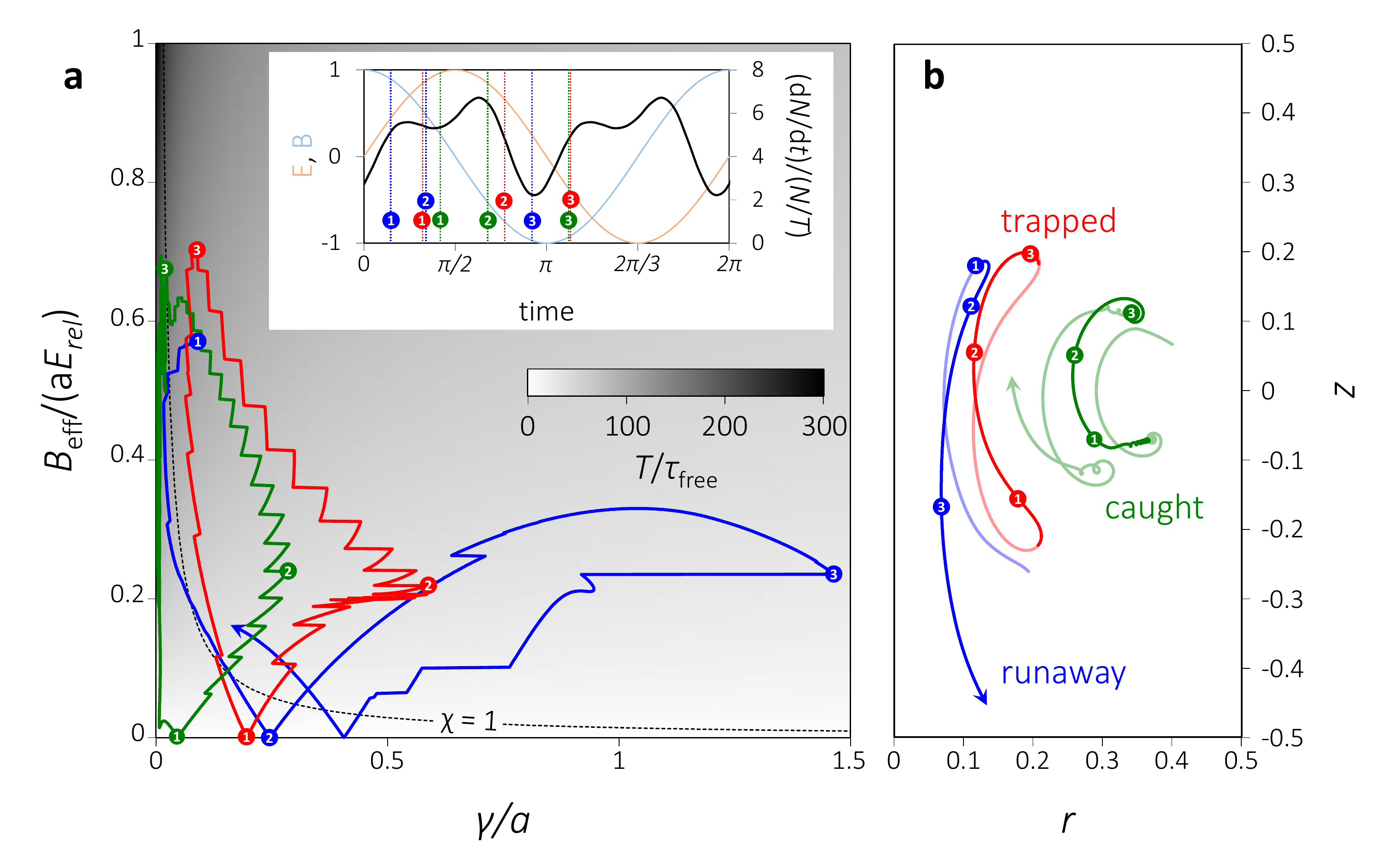}
	\caption{\label{trajectories} Representation of particle dynamics and emission for three possible regimes: migration to the trapped state after being caught (green), circulation in the trapped state (red) and leaving the trapped state (blue). The results obtained for $P = 40$~PW, using regularized (green and red) and stochastic (blue) QED emission models. Apart from the coordinate space (b), the dynamics is shown on the plane of $\gamma$ factor and the effective magnetic field $B_{\text{eff}}$ (a). The gray shades denote the mean free run time between emission events each of those one can be visible through a jump in the path pf the $B$-$\gamma$ plane. The insert indicates the phases of the electric (light red) and magnetic (light blue) fields and the relative increase of particles in the system due to the cascade as a function of time (black).}
\end{figure}

In Fig.~\ref{trajectories} we demonstrate how particles are caught (green) and circulate in the trapped state (red). The regularised emission model leads to particles sitting on stable attractors, a behaviour which in reality would break off due to stochastic effects. Hence in order to demonstrate how particles leave the system (blue) we use the now common probabilistic model of emission regularly employed in extended PIC schemes~\cite{gonoskov.pre.2015}. In panel (a) we show the paths of particles as a function of their $\gamma$-factor and the \textit{effective magnetic field} $B_{\text{eff}}$, which is defined as that field which would provide the instantaneous transverse acceleration of the particle at the particle's position (in space and time). These two parameters define the mean free motion time $t_{\text{free}}$ between emissions (gray-scale), while the average photon energy tends to roughly $1/4$ of the particle energy in the limit $\chi \gg 1$. The dark region in the left-upper of the figure denotes the state in which emission is frequent and thus energy gains are suppressed. Hence the optimal way for a particle to penetrate into the high-gamma region (in order to emit high-energy photons) is to pass through the low-$B$ region. As one can see, energy gain is suppressed by emission on the green trajectory, due to the strong magnetic field further from the $z$-axis. For the red trajectory, however, the particle gains more energy when bypassing the dark region through point (1), and then emits high-energy photons around point (2). This is because the magnetic field is weaker near the $z$-axis, as for the red trajectory (see panel (b)). In this way, the ART provides close-to-optimal particle dynamics as compared, for example, with the case of `normal radiative trapping'~\cite{ART}. That is why the dipole wave, apart from providing the highest possible electric field amplitude also enables the optimal strategy for particles to gain high energy while the magnetic field remains small. At the same time, ART causes particles to drift toward the optimal spatial region. Particles can occasionally emit very high energy photons, due to stochastic effects, and then leave the system: fortunately the cascade compensates this as it provides a self-sustained source of particles.

Pair production cascades are known to be complex processes\cite{grismayer.pop.2016, nerush.pop.2011, elkina.prstab.2011, bashmakov.pop.2014}. In our case this complexity can be seen in the black curve in the insert of Fig.~\ref{trajectories} (a); the characteristics of the particle growth rate vary over the entire wave period. This is due to the rate being a function of where particles are located in the $B$-$\gamma$, and this `position' varies greatly (note though that the particles are almost always in the region $\chi > 1$, as denoted by the dashed curve). In particular, the cascade is dominantly of avalanche type until the point \ding{173} (as the particles gain energy), while in the interval between points \ding{173} and \ding{174} the cascade is dominantly of shower type \cite{mironov.pra.2014}. These two types cannot be clearly distinguished or separated, though. Moreover, the contribution of a particle to the cascade depends strongly on its distance from the $z$-axis (i.e.~on the type of particle motion), while the spatial distribution of particles is determined by the stochasticity of emission and the rate of particle production in different states. This makes the system particularly intractable to simple analytic estimates, and hence an accurate analysis requires numerical simulations. However, particle motion, particle generation, and events of particles leaving the system are all governed by a `self-contained' process that is, as we will see below, defined only by the parameter $P$. That is why the presented analysis for all values of $0 < P < 100$~PW has a general relevance.

The simulations indicate that the cascade and ART appear in an interdependent way already at the power of just 7.2~PW, which is significantly lower than the 100~PW needed for the dominance of ART itself~\cite{ART}. Phenomenologically, this is because ART attracts both electrons and positrons to the high-field region, where the particles gain and emit more energy in the form of high-energy photons. This induces favourable conditions for the cascade. The cascade in turn serves as a source of particles already in the high-field region, where the trapping occurs for lower intensities than required for the dominance of ART in the other regions. In such a way, the cascade and ART induce and support each other.

\subsection{Estimate of the cascade threshold}

In the considered configuration, particles and generated photons propagate almost along the $z$-axis (i.e.~along the electric field vector). Thus $\chi$ and consequently the rates of photon emission and pair production are predominantly determined by the magnetic field strength. Hence in order to estimate the cascade onset threshold we assume that the cascade first arises in the instant when the magnetic field peaks, and in the spatial region where the particles and generated photons are in this instant of time, i.e. $z_{eff} \approx \pm \lambda / 4$, $r_{eff} \approx 0.15 \lambda$ (see Fig.~\ref{trajectories}~(b). As one can see from Fig.~4 of~reference\cite{ART}, $r_{eff}$ does not depend strongly on $P$). Using the analytical expressions for the dipole wave\cite{gonoskov.pra.2012} we can estimate that in this spatiotemporal region the effective magnetic field in relativistic units is $B_{eff} \approx 2.3 a r_{eff}/\lambda \approx 0.3 a$. For our estimate, we assume that a magnetic field of this strength acts for 1/8 of the wave period, which we associate with the typical duration of the turning phase. We also assume that particles and photons have an energy order of $mc^2 a$. We can then estimate $\chi \approx 0.3 a^2(mc^2/\hbar\omega)$. For the considered wavelength $\lambda = 910$~nm we have $\chi \approx 0.5P$. Thus for values of $P \gtrsim 10$~PW we can assume $\chi \gg 1$. In this case, we can neglect the time of photon generation, as it is roughly four times less than the one of pair production\cite{Ritus}, and estimate the threshold $P$ for the cascade to arise by requiring the average time of pair production to be equal to one eighth of the wave period. Using the large-$\chi$ approximation for pair production rate\cite{Ritus}, we obtain the relation
\be
0.37 \frac{\alpha m^2 c^4}{\hbar m c^2 a} \chi^{2/3} \frac{1}{8} \frac{2\pi}{\omega} = 1.
\ee
Using the relation $a \approx 800 \sqrt{P/(1 \text{PW}) }$ yields
\be
	P_\text{min} \approx 15\text{ PW} \;.
\ee

\subsection{Simulations}

Numerical simulation of the above described process requires accurate accounting for the radiation losses in the entire range of regimes from weak classical ($\chi \ll 1$) to strong quantum ($\chi \gg 1$). The former is important before the particles come close to the $z$-axis (point (3) of the green trajectory in fig.~\ref{trajectories}) or when the high field values have not been reached. The latter is important for simulating trapped and leaving particles. That is why for this study we developed a \textit{modified event generator}\cite{gonoskov.pre.2015} that reproduces exactly the particle emission without any low-energy cutoff (common for all previously known methods) for all values of $\chi$. Thus, in particular, it reproduces the classical emission in the limit $\chi \ll 1$. Despite of the fact that our method provides close to the minimal possible requirements to the time step, the interval between emissions and pair production can be extremely small in the central region of interaction. To bring the computational demands into realistic range we account for an arbitrary number of QED events within a single time-step using an \textit{adaptive event generator}\cite{gonoskov.pre.2015}. Additionally we apply procedures for resampling the ensemble of particles to avoid the memory overload through a gradual increase of the particles/photons computational weights. The validity of our model for the considered process is accurately demonstrated in the specially dedicated paper\cite{gonoskov.pre.2015}, where we thoughtfully analyzed all the methodological and algorithmical aspect and benchmarked the model against analytical results and other numerical codes. \\

For simulations we used our numerical model integrated as a module in a supercomputer 3D particle-in-cell code PICADOR \cite{bastrakov.jcp.2012}. For performing simulations we used the following setup. The simulation box of 4~$\mu m$ in each direction has been used together with a thin PML layer at the boundaries to mimic a part of infinite space by generating the incoming diploe wave and absorbing the outgoing emission. The considered target was a uniform plasma region of spherical shape with the diameter of 3~$\mu m$. The initial temperature was adjusted for different initial densities to make the Debye length resolvable by the grid. In all the simulations the initial temperature was below $10^{-3}$~eV and it's actual value did not affect the further dynamics. The time step of 20 as provided proper resolving of both plasma and field oscillations. \\

For determining the rate of particles number growth, we performed a set of simulations. To from an $e$-dipole wave with a constant amplitude we implemented inwards generation (within the PML layer at the boundaries) of an inverse emission of a dipole antenna (see \cite{gonoskov.pra.2012}) with the wavelength of 0.9~$\mu m$ and a smoothed temporal transition of the amplitude from zero to a constant value, which has been varied. The transition had a form of an inverse tangent and the length of two cycles. The number of particles in the central region was recorded as a function of time. Based on this the average rate of the exponential growth was determined. All the outlined particular conditions have been checked to not affect the obtained values for the rate. \\

For the numerical experiments with realistic laser pulses we used a Gaussian temporal profile for the incoming dipole wave with a duration of 15~fs (FWHM for intensity). For the considered total peak power of 40~PW the initial density of $10^{13}$~cm$^{-3}$ provided the absence of self-action within the entire duration of the numerical experiment. \\

\subsection{Analytical estimates and calculations}

\begin{figure}[t!]
	\centering
	\includegraphics[width = 0.45\columnwidth]{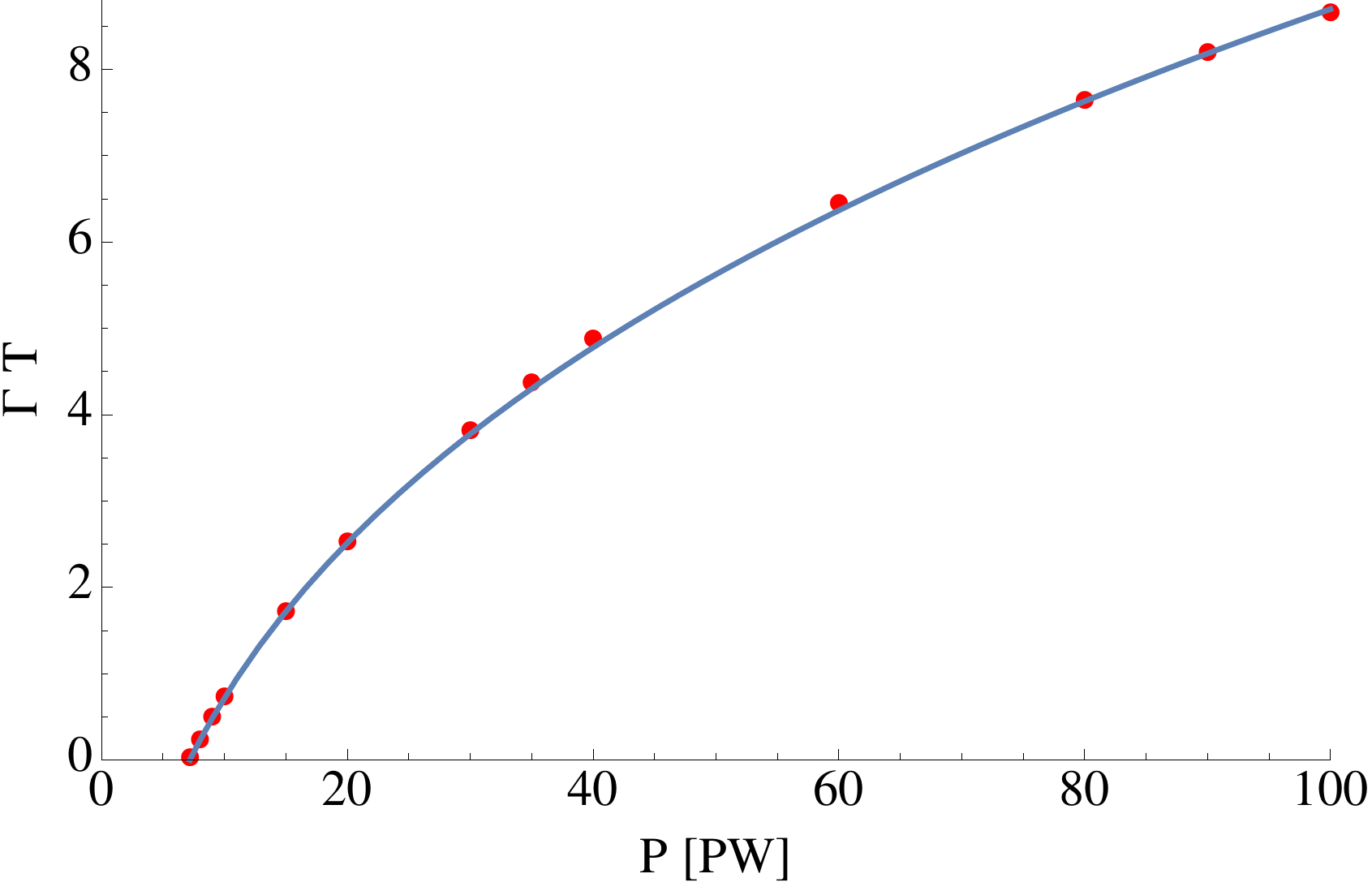}
	\caption{The measured averaged rate of the exponential growth for the number of particles produced and trapped in the center of a dipole wave as a function of the total power P. The red dots indicate numerical experiments and the dashed line shows the approximation used in the further analysis.
	}
	\label{fig_rate}
\end{figure}

The cascade development within each turning phase is not exponential. However, because the produced particles are trapped by the ART mechanism they generate very similar cascades during the next turning phase, and as a result the number of particles does grows exponentially on timescales longer than a half-period. Let $\Gamma$ be the average rate over a half-period; simulation results for $\Gamma$ multiplied by the laser period $T$ are shown in Fig.~\ref{fig_rate} as a function of total power $P$. In order to provide some further analytic estimates we model this curve by the solid line shown in the figure, which has the form
\be
	\Gamma(P) T =  \kappa\big( P^{1/3} - P_\text{min}^{1/3}\big) \;,
\ee
where $P_\text{min} \simeq 7.2$~PW is the threshold value required for the cascade to be initiated, and $\kappa= 3.21$, both parameters numerically determined.

\begin{figure}[t!]
	\includegraphics[width = 0.45\columnwidth]{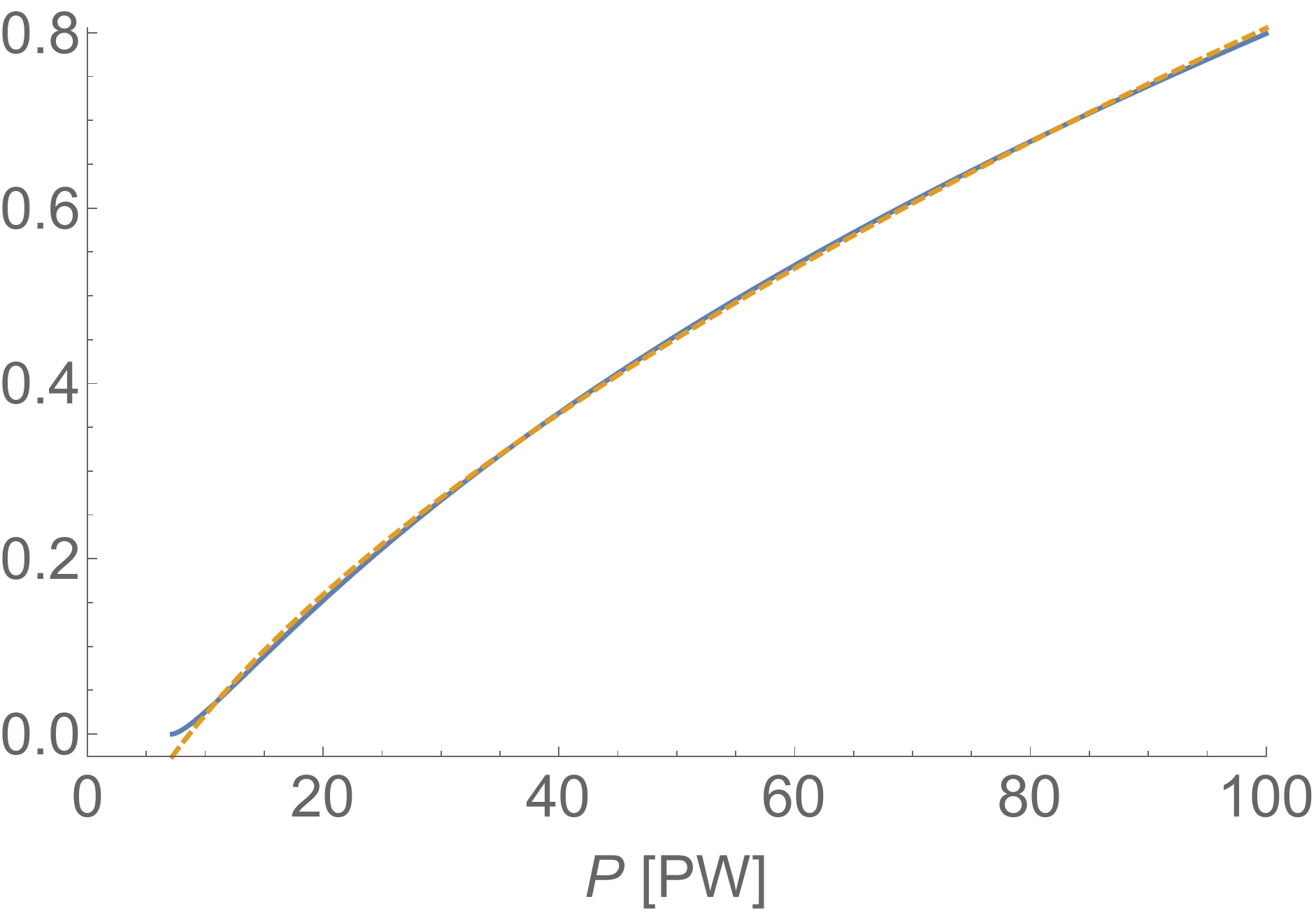}
	\caption{\label{FIG:JAEMN} Comparison of the exact and approximate expression for $f(P/P_\text{min})$ from equations (\ref{F1}) and (\ref{METHODSest2}). }
\end{figure}

The equation describing the growth of particle number $N(t)$ is:  
\begin{equation}\label{METHODSgrowth1}
	\frac{\partial }{\partial t} N(t) = \Gamma\big(P(t)\big) N(t) \;.
\end{equation}
Assuming that the dipole wave has a Gaussian temporal shape with peak power $P_0$ and duration $\tau$ (FWHM for the power), so that
$P(t) =P_0 e^{-4\ln(2) t^2/\tau^2}$, we can integrate (\ref{METHODSgrowth1}) and a relation between the growth of the particle number from an initial value $N_0$ to $N_{\text{max}}$ at the instance of peak intensity, $t = 0$. Writing $x\equiv P/P_\text{min}$ we find
\begin{equation}\label{METHODSest1}
	\log \frac{N_\text{max}}{N_0} = \frac{\tau_\text{opt}}{T} \frac{\kappa P_\text{min}^{1/3}\sqrt{3}}{\sqrt{4\log 2}} f\big(P/P_\text{min}\big)
\end{equation}
where
\be\label{F1}
	f(x)=x^{1/3} \frac{\sqrt{\pi}}{2}\text{Erf}(\sqrt{\log x^{1/3}})- \sqrt{\log x^{1/3}}. 
\ee
For $P$ in the range $P_\text{min}\ldots 100$ the Error function may be approximated by its series expansion, but the first three nontrivial orders are needed to give good agreement over the whole range. The asymptotic expansion does not seem to work over this range. By rather crudely fitting points we can approximate, for $P$ above threshold and less than $100~PW$,
\begin{equation}\label{METHODSest2}
	f(P/P_\text{min}) \sim 0.07 (P^{0.59} - 3.6) \;,
\end{equation}
which gives a good approximation over the whole range, the only real deviation being at the threshold, see Fig.~\ref{FIG:JAEMN}.

To proceed we need estimates for $N_0$ and $N_\text{max}$. As these appear inside the logarithm in (\ref{METHODSest1}) the estimates can be quite rough. We first choose $N_0$ large enough to be safely sure that some particle generates cascades: $N_0 = 10$, is sufficient. In order to identify the upper limit $N_\text{max}$ we equating the Coulomb field at a distance of $\rho_{\text{eff}}$ to $1/10$ of the peak field of the dipole wave (i.e.~$10\%$ self-interaction effects): this yields $N_{\text{max}} \approx 10^9\sqrt{P_0}$, which as we will see below fits well with simulation results.

\be
	\frac{\tau}{T} \simeq \frac{2.23 \log[10^8 P^{1/2}] } {P^{0.59}-3.6}
\ee

\subsection{Numerical calculation of the photon yield for a Gaussian pulse}

To confirm that the proposed concept is optimal for producing a maximal number of photons we use the results of numerical experiments with constant power $P$ to calculate the number of photons that would be generated with an arbitrary Gaussian pulse. We use the following procedure. 

\begin{figure}[t!]
	\includegraphics[width = 0.9\columnwidth]{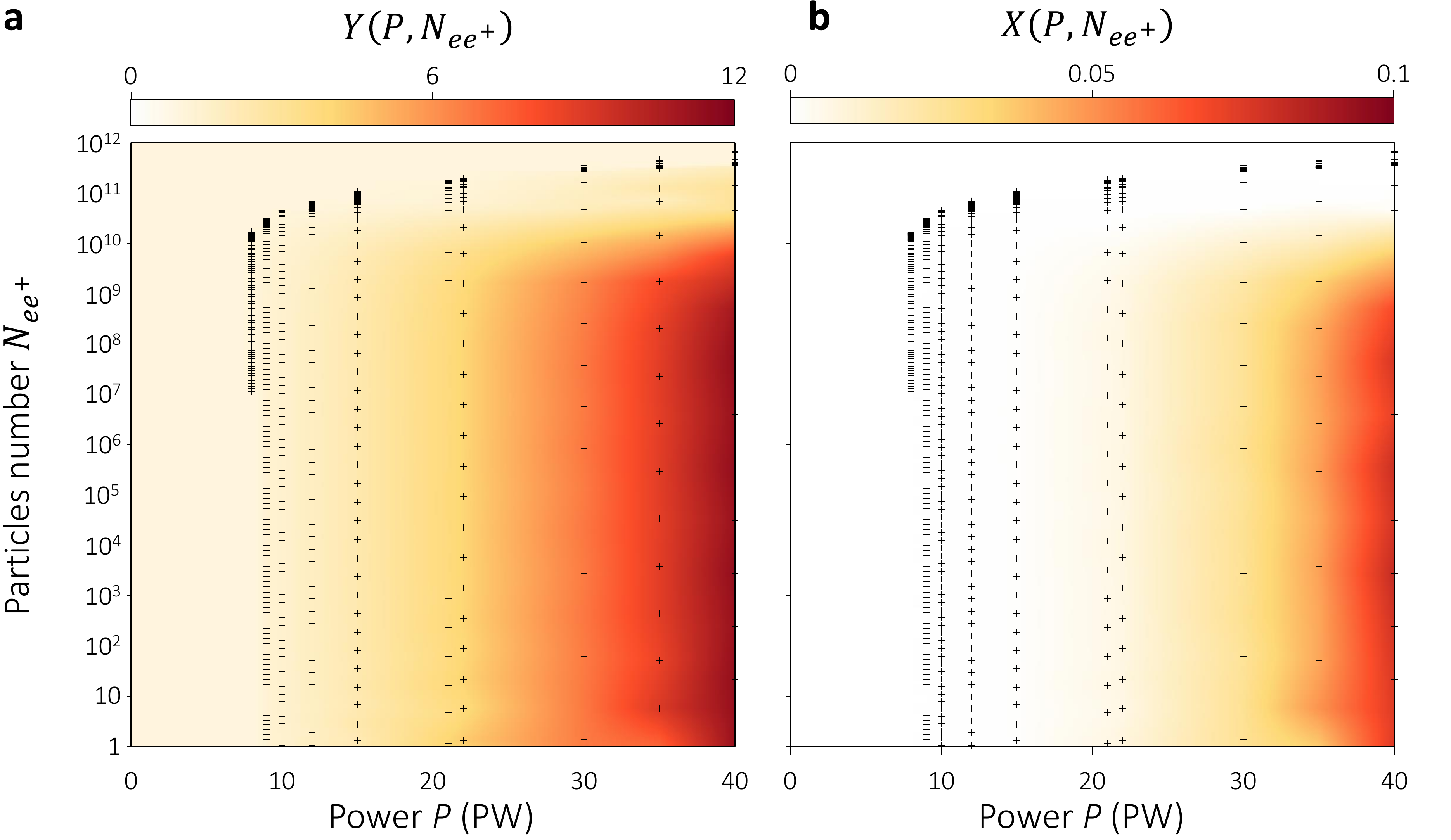}
	\caption{The functions $X\left(P, N_{ee^+}\right)$ and $Y\left(P, N_{ee^+}\right)$ interpolated from the values calculated from the data of simulations with constant power $P$ at the instances of $E = 0$ (crosses).}
	\label{FIG:XY}
\end{figure}

All intervals between instances of $E = 0$ are, despite the complexity of the evolution within each half-cycle, self similar. This is because the particles inherit almost no information from previous half-cycles after loosing energy due to strong magnetic field (when it peaks). Thus, assuming a slow variation of the wave amplitude, we can describe the discreet evolution of the particles number $N_{ee^+}$ and the number of generated photons $N_{ph}$ (which escaped from the strong field region) at the instances of $E = 0$ using the following equations:
\be
\label{evolution_XY}
\begin{array}{lll}
	N_{ee^+} \left(t + T/2\right) = N_{ee^+}\left(t\right) Y\left(P\left(t + T/4\right), N_{ee^+}\left(t\right)\right),\\
	N_{ph}\left(t + T/2\right) = N_{ph}\left(t\right) + N_{ee^+}\left(t\right) X\left(P\left(t + T/4\right), N_{ee^+}\left(t\right)\right),
\end{array}
\ee
where the functions $X\left(P, N_{ee^+}\right)$ and $Y\left(P, N_{ee^+}\right)$ are defined as the relative increase in the number of particles and the number of generated photons within a single half-period between instances of $E = 0$, for the current value of power $P$ and number of particles $N$ at the beginning of the interval. For our study we assume that $N_{ph}$ is the number of photons with energy above 2~GeV. We use the data obtained in the set of experiments with constant power $P$ and interpolate the values in other points. The result is presented on the fig.~\ref{FIG:XY}. Using equations (\ref{evolution_XY}) and the numerically determined functions $X\left(P, N_{ee^+}\right)$ and $Y\left(P, N_{ee^+}\right)$ we then can calculate the expected number of photons to be generated by an arbitrary Gaussian pulse. The results are presented on the fig.~\ref{fig:Fig2}(b).

\section*{Acknowledgements}

The authors acknowledge support from the Russian Science Foundation project No.~16-12-10486 (analytical part of the work, A.G., A.B., E.E.,A.M.,A.K.), the Russian Foundation for Basic Research projects No. 15-37-21015 (numerical experiments, A.G., A.M, S.B., I.M.), a Marie-Curie Individual Fellowship, project 701676  (A.I.) the Olle-Engkvist Foundation grant 2014/744 (A.I.), the Knut and Alice Wallenberg  Foundation (A.G., A.I., M.M.). The simulations were performed on resources provided by the Swedish National Infrastructure for Computing (SNIC) and the Joint Supercomputer Center of RAS. A.G. and A.B. contributed equally.

\end{document}